\documentclass{JHEP3}
\usepackage[dvips]{graphicx} 

\title{A consistent description of $\rho^0\rightarrow\pi\pi\gamma$ decays 
including $\sigma(500)$ meson effects}
\author{
Albert Bramon\\
Grup de F\'{\i}sica Te\`orica,
Universitat Aut\`onoma de Barcelona,
E-08193 Bellaterra (Barcelona), Spain\\
E-mail: \email{bramon@ifae.es}
}
\author{
Rafel Escribano\\
Grup de F\'{\i}sica Te\`orica and IFAE,
Universitat Aut\`onoma de Barcelona,
E-08193 Bellaterra (Barcelona), Spain\\
E-mail: \email{Rafel.Escribano@ifae.es}
}
\abstract{
A consistent description of $\sigma(500)$ meson effects in
$\rho^0\rightarrow\pi^0\pi^0\gamma$ and $\pi^+\pi^-\gamma$ decays is 
proposed in terms of reasonably simple amplitudes which reproduce the expected
chiral-loop behaviour for large $m_{\sigma}$ values.  
For the neutral case, in addition to the well known $\omega$ exchange,
there is an important contribution from the $\sigma(500)$ meson that is in
agreement with recent experimental data.
For the charged case, where the dominant contribution comes from bremsstrahlung, 
the effects of the $\sigma(500)$ meson are relevant only at high 
values of the photon energy and compatible with present data.
A combined analysis of both processes with moderately improved experimental information
should contribute decisively to clarify the status of this controversial 
$\sigma(500)$ meson.
}
\keywords{sig, pmo, chl}
\preprint{UAB--FT--545}

\begin{document}
\section{Introduction}
If there is a meson resonance whose existence or not is still an
open question in spite of many dedicated discussions,  
this is the $\sigma(500)$ meson. Although the current PDG edition 
\cite{Hagiwara:fs} classifies this
scalar state ---the  $\sigma$ or $f_0$(400--1200)--- among the stablished resonances,
this has not been the case for most of a controversial period starting some 30 years ago. 
Data on $\pi \pi$ scattering at low energies, whose isoscalar $s$-wave channel
should reflect the $\sigma(500)$ effects and allow for the extraction of 
the $\sigma(500)$ properties, 
have resisted unambiguous analyses. Only recently, a growing number of authors claim for
the existence of such a $\pi \pi$ resonant state with a mass around some 500 MeV and a
similar width (for two recent reviews,
see Refs.~\cite{Tornqvist:2002es} and \cite{vanBeveren:2002vw}).
But the controversy on the existence of the $\sigma(500)$, as well as on its nature and
properties, is still open.
The purpose of our note is to illustrate that a combined analysis of the radiative 
$\rho^0\rightarrow\pi^0\pi^0\gamma$ and $\rho^0\rightarrow\pi^+\pi^-\gamma$ decays should
considerably contribute to clarify the issue. 

The contribution of the $\sigma(500)$ meson to the amplitudes of these two radiative
processes is exactly the same under isospin invariance,  
${\cal A}(s)_\sigma \equiv {\cal A}(\rho\rightarrow\pi^0\pi^0\gamma)_\sigma=
 {\cal A}(\rho\rightarrow\pi^+\pi^-\gamma)_\sigma$.
It should be the dominant one in
the $s$-channel ($s \equiv m^2_{\pi\pi} \leq m^2_{\rho}$) where two-pion resonance formation
with $J^{PC} = 0^{++}, 2^{++}$\ldots\ can occur. Indeed, while the $\sigma(500)$ has 
a mass below $m_{\rho}$ and strongly couples to pion pairs, other resonant exchanges have
too large a mass (like the $f_2 (1270)$) or almost decouple from pions (like the $f_0(980)$).
This common $\sigma$ amplitude 
---the {\it signal} amplitude, ${\cal A}(s)_\sigma$--- will interfere with
other contributions ---{\it background} amplitudes--- accounting for other exchanges.
The latter can be reliably computed for both the neutral and 
charged decays and turn out to be markedly different. Thanks to this, the  
combined study of both decays and comparison with their data considerably constraints the 
common signal amplitude, ${\cal A}(s)_\sigma$, and should allow for the 
extraction of the $\sigma(500)$ meson properties. 
With the available data on $\rho^0\rightarrow\pi^0\pi^0\gamma$ 
\cite{Achasov:2002jv,Achasov:2000zr} 
and $\rho^0\rightarrow\pi^+\pi^-\gamma$ \cite{Dolinsky:vq,Vasserman:yr} one 
already can infer that a low-mass $\sigma(500)$ resonance is most likely 
required. More accurate data coming from the Frascati $\phi$-factory 
DA$\Phi$NE \cite{daphne95:franzini} could confirm this conclusion and extract the relevant 
$\sigma(500)$ meson properties.

\section{The common, $\sigma(500)$-dominated amplitude}
The suitable tool to study $e^+ e^-$ annihilation into 
$\pi^0\pi^0\gamma$ or $\pi^+\pi^-\gamma$ well below the $\rho$ resonance pole 
is Chiral Perturbation Theory ($\chi$PT) \cite{Gasser:1984gg}.
At the one-loop level, this would imply the
extension of the analysis on $\gamma\gamma\rightarrow\pi^0\pi^0$ 
and $\gamma\gamma\rightarrow\pi^+\pi^-$ for real photons performed in 
Ref.~\cite{Bijnens:1987dc} to the case where one photon is off mass-shell
$(q^*)^2 \neq 0$. But if the $\rho$ resonance pole is approached, $(q^*)^2 \simeq m^2_\rho$, 
$\chi$PT no longer applies and one needs to enlarge this theory to include resonances. There
is some consensus in that vector and axial-vector mesons have to be incorporated in such a
way that the old and successful ideas of Vector Meson Dominance (VMD) are fulfilled 
\cite{Ecker:1988te}, but the situation concerning the inclusion of scalars is notoriously more 
ambiguous \cite{Ecker:1988te,Bramon:1994bw}.  

Extensions of $\chi$PT including vector mesons which are suitable for the analysis of 
$\rho^0\rightarrow\pi\pi\gamma$ radiative decays have been presented elsewhere 
\cite{Bramon:1992kr,Bramon:1992ki}.
They are particularly simple when specified to the neutral decay mode
$\rho^0\rightarrow\pi^0\pi^0\gamma$. In this case one has to compute the same set of
one-loop diagrams contributing to $\gamma\gamma \rightarrow \pi^0\pi^0$ shown in
Ref.~\cite{Bijnens:1987dc}, the only difference being the substitution of one photon with 
$(q^*)^2 \neq 0$ by a massive $\rho$ meson according to VMD.
Restricting to the contribution from charged-pion loops (charged-kaon loops contribute negligibly 
\cite{Bijnens:1987dc,Bramon:1992ki}) one obtains the following finite amplitude for 
$\rho(q^\ast,\epsilon^\ast)\rightarrow\pi^0(p)\pi^0(p^\prime)\gamma(q,\epsilon)$: 
\begin{equation} 
\label{Achiralneutral} 
{\cal A}(\rho\rightarrow\pi^0\pi^0\gamma)_\chi=  
\frac{-eg}{\sqrt{2}\pi^2 m^2_{\pi^+}}\,\{a\}\,L(m^2_{\pi^0\pi^0})\times 
{\cal A}(\pi^+\pi^-\rightarrow\pi^0\pi^0)_\chi\ , 
\end{equation}  
where 
$\{a\} \equiv (\epsilon^\ast\cdot\epsilon)\,(q^\ast\cdot q)- 
       (\epsilon^\ast\cdot q)\,(\epsilon\cdot q^\ast)$,   
$m^2_{\pi^0\pi^0}\equiv s\equiv (p+p^\prime)^2=(q^\ast -q)^2$ is the invariant mass of  
the final dipion system and $L(m^2_{\pi^0\pi^0})$ is the loop integral function  
defined in Refs.~\cite{Bramon:1992ki}--\cite{Close:1993ay}.
The coupling constant $g$ comes from the strong amplitude 
${\cal A}(\rho\rightarrow\pi^+\pi^-)=-\sqrt{2}g\,\epsilon^\ast\cdot (p_+-p_-)$  
with $|g|=4.24$ to agree with $\Gamma (\rho\rightarrow\pi^+\pi^-)_{\rm exp}=149.2$ MeV
\cite{Hagiwara:fs}. 
The final factor in Eq.~(\ref{Achiralneutral}) is 
\begin{equation} 
\label{A4Pchiralneutral} 
{\cal A}(\pi^+\pi^-\rightarrow\pi^0\pi^0)_\chi=\frac{s - m^2_\pi}{f_\pi^2}\ . 
\end{equation}
Although it is the part of the amplitude which is potentially sensitive to the effects of 
$\sigma$ resonance formation, this is not contemplated in our chiral-loop evaluation at lowest
order.
By itself this ${\cal A}(\pi^+\pi^-\rightarrow\pi^0\pi^0)_\chi$ amplitude
in Eq.~(\ref{Achiralneutral}) ---devoid of $\sigma$ formation effects--- leads to 
\begin{equation} 
\label{Gchiralneutral} 
\Gamma (\rho\rightarrow\pi^0\pi^0\gamma)_\chi=1.55\ \mbox{keV}\ , 
\end{equation} 
for $f_{\pi}=92.4$ MeV.
It is worth mentioning that the amplitude (\ref{Achiralneutral}) is 
calculated by means of the ${\cal O}(p^2)$ $\chi$PT Lagrangian, ${\cal L}_{2}$,
enlarged to include external vector meson fields through the covariant 
derivative.
In this sense, the $\pi^+\pi^-\to \pi^0\pi^0$ amplitude in Eq.~(\ref{A4Pchiralneutral})
is correct only at lowest order in the chiral expansion.
A more refined two loop analysis including terms of the
${\cal O}(p^4)$ $\chi$PT Lagrangian, ${\cal L}_{4}$, would make the amplitude 
(\ref{A4Pchiralneutral}) no longer proportional to 
$(s-m_{\pi}^2)$ but corrected by chiral loop and counterterm 
contributions \cite{Gasser:1983yg,GomezNicola:2001as}.
Some of these counterterms are known to contain the effects of scalar resonance 
exchange \cite{Ecker:1988te}.
If one is only interested in such effects, as in our present case,
it has been shown very recently that a direct comparison of 
$\pi\pi$ scattering in the Linear Sigma Model (L$\sigma$M) and $\chi$PT at ${\cal O}(p^4)$
fixes the relevant counterterms in such a way that the $\pi^+\pi^-\to \pi^0\pi^0$ 
amplitude is still proportional to $(s-m_{\pi}^2)$ \cite{Bramon:2003xq}.
However, the advantage of using a framework where the scalar 
resonances are taken into account explicitly is that it allows to reproduce 
the scalar pole effects, a feature that is not possible in $\chi$PT.

As stated, the $\sigma$ resonance formation effects should modify the previous results.
In particular, instead of Eq.~(\ref{A4Pchiralneutral}) one now has to expect 
\begin{equation} 
\label{A4PchiralneutralF} 
{\cal A}(\pi^+\pi^-\rightarrow\pi^0\pi^0)_F =\frac{s - m^2_\pi}{f_\pi^2} 
F_\sigma (s) \ , 
\end{equation}
with  an additional factor 
\begin{equation} 
\label{FF}
F_\sigma (s) \equiv {-m^2_\sigma  +k m^2_\pi \over D_\sigma (s)} \ ,
\end{equation}
accounting for $\sigma$ exchange. 
A simple Breit-Wigner form,
$D_\sigma (s) \equiv s-m^2_\sigma +i m_\sigma \Gamma_\sigma$,
where $m_{\sigma}$ and $\Gamma_\sigma$ are the effective mass and width,
will be assumed for the $\sigma$ propagator and
contributions from $f_0 (980)$ exchange will be neglected (they can be estimated
to be below some 2 per thousand). Note that, as required, $F_\sigma (s) \to 1$ 
when $m^2_\sigma \to \infty$ for any finite value of a free parameter 
$k$,  thus recovering the chiral-loop result in Eq.~(\ref{A4Pchiralneutral}).
It is important to remark that in Eq.~(\ref{A4PchiralneutralF}) we are not adding the $\sigma$ 
contribution \textit{ad hoc} but in a way that preserves the lowest order $\chi$PT amplitude once
the $\sigma$ resonance is decoupled.
As mentioned before, the $\pi^+\pi^-\to \pi^0\pi^0$ amplitude is 
shown to be proportional to $(s-m_{\pi}^2)$ even when the $\sigma$ 
formation effects are taken into account.
This feature together with the recovery of the chiral result 
makes of the amplitude (\ref{A4PchiralneutralF}) a valid amplitude for 
studying such effects, thus making the whole analysis quite reliable.

We will consider two possible values
of the parameter $k$: $k = 1$  and $k \simeq -2.5$. The first value corresponds to the Linear
Sigma Model (L$\sigma$M) \cite{Tornqvist:1999tn}--\cite{Escribano:2002aj}
for scalar resonances, where the $\sigma\pi\pi$ coupling is given by 
$g_{\sigma\pi\pi} = (- m^2_\sigma + m^2_\pi )/f_\pi$. Once inserted in 
\begin{equation} 
\label{swidth}
\Gamma_{\sigma} \simeq \Gamma (\sigma \to \pi \pi)  =
{3\over 32\pi} {g^2_{\sigma\pi\pi} \over m_\sigma} \sqrt{1- {4 m^2_\pi \over m^2_\sigma}}\ , 
\end{equation}
it predicts a $\sigma$ (total) width around 300 MeV, which is only slightly below the 
value $\Gamma_{\sigma} \simeq 500$ MeV favoured in 
Refs.~\cite{Tornqvist:2002es,vanBeveren:2002vw}. 
This favoured value is reproduced if we enlarge the $g_{\sigma\pi\pi}$ coupling constant 
by fixing instead 
$k \simeq -2.5$. By itself, the amplitude for each one of these values of $k$ ($k=1$ in the
L$\sigma$M or $k \simeq -2.5$ in a more phenomenological context) inserted as the final factor in 
Eq.~(\ref{Achiralneutral}) predicts, respectively,  
\begin{equation} 
\label{GchiralneutralF} 
\Gamma (\rho\rightarrow\pi^0\pi^0\gamma)_{\mbox{\scriptsize L$\sigma$M}}=2.63\ \mbox{keV}\ ,
\quad
\Gamma (\rho\rightarrow\pi^0\pi^0\gamma)_{\mbox{\scriptsize $\sigma$-phen}}=1.84\ \mbox{keV}\ . 
\end{equation}
The differences among the results in Eqs.~(\ref{Gchiralneutral}) and 
(\ref{GchiralneutralF}) illustrate the effects of the $\sigma$(500) resonance 
in  $\rho^0\rightarrow\pi^0\pi^0\gamma$ decays and seem to be large enough to establish 
both its existence and total width.

The same set of diagrams as before \cite{Bijnens:1987dc} contributes 
(apart from another set to be discussed later) to the amplitude for the charged
channel  
$\rho^0\rightarrow\pi^+\pi^-\gamma$. 
It similarly leads to  
\begin{equation} 
\label{Achiralcharged} 
{\cal A}(\rho\rightarrow\pi^+\pi^-\gamma)_\chi=  
\frac{-eg}{\sqrt{2}\pi^2 m^2_{\pi^+}}\,\{a\}\,L(m^2_{\pi^+\pi^-})\times 
{\cal A}(\pi^+\pi^-\rightarrow\pi^+\pi^-)_\chi\ , 
\end{equation}  
where the  
four-pseudoscalar amplitude factorizes again in Eq.~(\ref{Achiralcharged}) 
but now it is found to be 
proportional to the variable $s=m^2_{\pi^+\pi^-}=m^2_{\rho}-2m_{\rho}E_{\gamma}$: 
\begin{equation} 
\label{A4Pchiralcharged} 
{\cal A}(\pi^+\pi^-\rightarrow\pi^+\pi^-)_\chi=\frac{s}{2f_\pi^2}\ . 
\end{equation} 
Integrating the photon energy spectrum over the whole physical region as before, one obtains 
\begin{equation} 
\label{Gchiralcharged}
\Gamma(\rho\rightarrow\pi^+\pi^-\gamma)_{\chi}=0.93\ \mbox{keV}\ , 
\end{equation} 
which is the simple chiral-loop prediction with no $\sigma$ meson effects. These are easily
introduced in terms of the previous $F_\sigma (s)$ factor accounting for  $\sigma$ resonance
formation in the $s$-channel. Since isospin invariance forces this 
$\sigma$ contribution to coincide with that for the
previous neutral case, one unambiguously has 
\begin{equation}
\label{A4PchiralchargedlF} 
{\cal A} (\pi^+\pi^-\rightarrow\pi^+\pi^-)_F=\frac{s - m^2_\pi}{f_\pi^2} 
F_\sigma (s) - \frac{s/2 - m^2_\pi}{f_\pi^2}  \ .
\end{equation}
The final term is hard to interpret physically but it cannot be associated to $\sigma$ formation
in the $s$-channel and, as such, it does not contain the $F_\sigma (s)$ factor. 
It could be understood as the contribution of the exchange of all other 
intermediate resonances in the infinite mass limit.
In any case, it is totally fixed by the need to recover the chiral-loop result 
(\ref{A4Pchiralcharged}) in the limit  
$m^2_\sigma \to \infty$ or $F_\sigma (s) \to 1$. 
Once inserted as the final factor in Eq.~(\ref{Achiralcharged}), 
the two terms in amplitude (\ref{A4PchiralchargedlF}) lead to
${\cal A}(s)_\sigma +
 {\cal A}(\rho\rightarrow\pi^+\pi^-\gamma)_{\mbox{\scriptsize non-$\sigma$}}$, 
with a first $k$-dependent term accounting for $\sigma$ exchange in the $s$-channel 
(as in the neutral case) and a second $k$-independent one.
For $k=1$ (as in the L$\sigma$M) or $k\simeq -2.5$ (as in the previous phenomenological context) 
one obtains  
\begin{equation} 
\label{GchiralchargedF} 
\Gamma (\rho\rightarrow\pi^+\pi^-\gamma)_{\mbox{\scriptsize L$\sigma$M}}=5.21\  \mbox{keV}\ ,
\quad
\Gamma (\rho\rightarrow\pi^+\pi^-\gamma)_{\mbox{\scriptsize $\sigma$-phen}}=3.84\ \mbox{keV}\ , 
\end{equation}
which, again, are markedly different from the value (\ref{Gchiralcharged}).

\FIGURE{
\centerline{\includegraphics[width=0.85\textwidth]{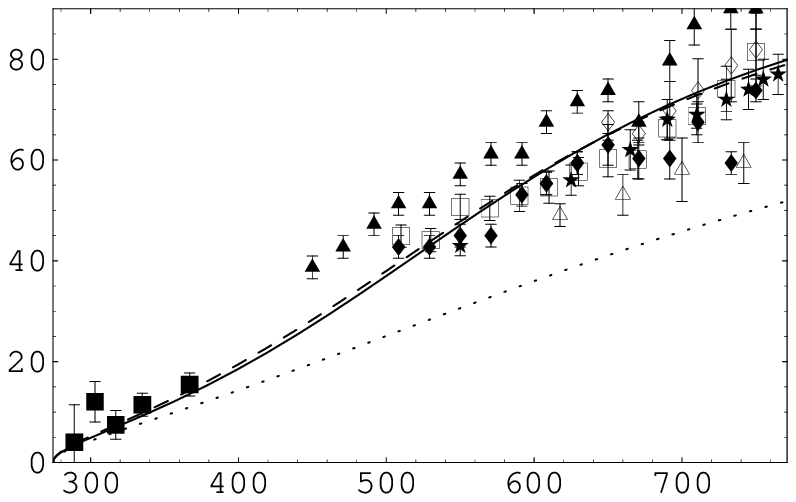}}  
\caption{\small 
$\delta^0_{0}(s)$ (degrees) 
as a function of the dipion invariant mass $m_{\pi\pi}$ (MeV). 
The various predictions are for the $\sigma$ models with $k=1$ (solid 
line), $k=-2.5$ (dashed line), and for the lowest order $\chi$PT (dotted line).
Experimental data are taken from different analyses of the CERN-Munich 
Collaboration \protect\cite{Grayer:1974cr} (open and solid diamonds and triangles),
as well as from \protect\cite{Protopopescu:1973sh} (stars),
\protect\cite{Rosselet:1976pu} (solid squares), and
\protect\cite{Estabrooks:1974vu} (open squares).} 
\label{plotphaseshift} 
}

The predictions for $\rho\rightarrow\pi\pi\gamma$ in both channels are 
thus clearly different if one takes into 
consideration the effects of $\sigma$ formation or not.
One would expect that this difference to be also manifest in $\pi\pi$ 
scattering itself.
In particular, one can calculate the effects of the $\sigma$ 
resonance for the $I=J=0$ $\pi\pi$ phaseshift $\delta^0_{0}(s)$
within the models with $k=1$ (L$\sigma$M) and $k=-2.5$ 
(phenomenological) 
and compare them with $\chi$PT at lowest order,
\textit{i.e.}~with no $\sigma$ resonance effects.
Following Ref.~\cite{Achasov:iu},
$\delta^0_{0}(s)=\sqrt{1-4m_{\pi}^2/s}\,T^0_{0}(s)$
where $T^0_{0}$ is the partial wave with $I=J=0$ obtained from the
$\pi^+\pi^-\to \pi^0\pi^0$ amplitude (see also Ref.~\cite{Black:2000qq}).
A comparison of the different models with experimental data is shown in
Fig.~\ref{plotphaseshift}.
Again, the models including $\sigma$ meson effects offer a better description
of data than the lowest order chiral prediction.
In both $\sigma$ models, with $k=1$ and $k=-2.5$, a best fit to the data is 
achieved for the effective parameters $m_{\sigma}\simeq 500$ MeV and
$\Gamma_{\sigma}\simeq 500$ MeV, in agreement with
Refs.~\cite{Tornqvist:2002es,vanBeveren:2002vw}.

\section{$\rho^0\rightarrow\pi^0\pi^0\gamma$}
Apart from the previously discussed amplitude, the $\rho^0\rightarrow\pi^0\pi^0\gamma$ decay is
known to proceed also via $\omega$-meson exchange in the $t$ and $u$ channels. Its evaluation
offers no problems and has been performed by many authors with coincident 
results 
(see, for instance \cite{Singer}). Explicitly, this background 
amplitude reads  
\begin{equation} 
\label{AVMD}
\textstyle
{\cal A}(\rho\rightarrow\pi^0\pi^0\gamma)_{\omega}=\frac{G^2 e}{\sqrt{2}g} 
\left(\frac{P^2\{a\}+\{b(P)\}}{M^2_\omega-P^2-i M_\omega\Gamma_\omega}+ 
      \frac{{P^\prime}^2\{a\}+\{b({P^\prime})\}} 
           {M^2_\omega-{P^\prime}^2-i M_\omega\Gamma_\omega}\right)\ , 
\end{equation} 
with $\{a\}$ the same as in Eq.~(\ref{Achiralcharged}) and 
\begin{equation} 
\label{b} 
\begin{array}{rl} 
\{b(P)\}\equiv & -(\epsilon^\ast\cdot\epsilon)\,(q^\ast\cdot P)\,(q\cdot P) 
           -(\epsilon^\ast\cdot P)\,(\epsilon\cdot P)\,(q^\ast\cdot q)\\[1ex] 
         & +(\epsilon^\ast\cdot q)\,(\epsilon\cdot P)\,(q^\ast\cdot P) 
           +(\epsilon\cdot q^\ast)\,(\epsilon^\ast\cdot P)\,(q\cdot P)\ , 
\end{array} 
\end{equation} 
where $P=p+q$ and $P^\prime=p^\prime+q$ are the momenta of the intermediate  
$\omega$ meson in the $t$- and $u$-channel, respectively, and $G$ is the strong 
and well known $\rho\omega\pi$ coupling constant \cite{Achasov:2002jv,Bramon:1992kr}. 
\FIGURE{
\centerline{\includegraphics[width=0.85\textwidth]{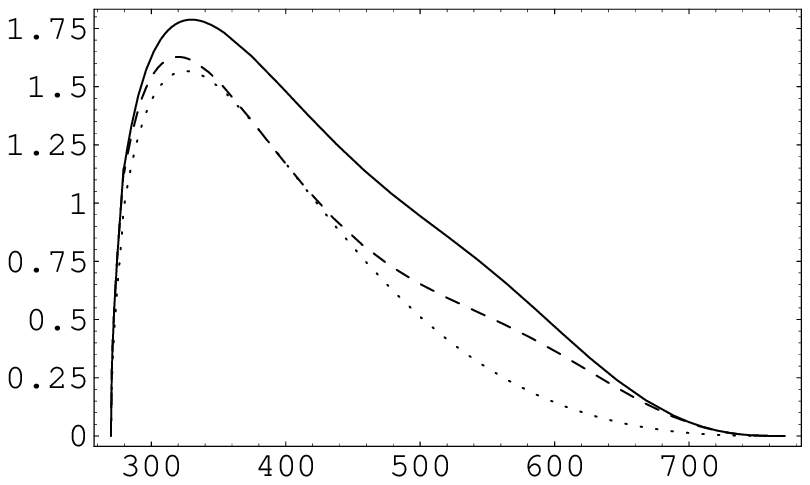}}  
\caption{\small 
$dB(\rho\rightarrow\pi^0\pi^0\gamma)/dm_{\pi^0\pi^0}\times 10^7\ (\mbox{MeV}^{-1})$ 
as a function of the dipion invariant mass $m_{\pi^0\pi^0}$ (MeV). 
The various predictions are for the input values:  
$m_\sigma = 500$ MeV and $\Gamma_\sigma = 300$ MeV (solid line);   
$m_\sigma = 500$ MeV and $\Gamma_\sigma = 500$ MeV (dashed line).
The chiral-loop prediction with no scalar pole is also included for comparison
(dotted line).} 
\label{plotneutral} 
}

From this VMD amplitude alone and $G=\frac{3g^2}{4\pi^2 f_{\pi}}$ one easily obtains    
\begin{equation} 
\label{GVMD} 
\Gamma(\rho\rightarrow\pi^0\pi^0\gamma)_\omega=1.89\ \mbox{keV}\ ,
\end{equation} 
in agreement with the results in Refs.~\cite{Achasov:2002jv,Bramon:1992kr}, once the 
slight  differences in numerical inputs are unified. The interference of this 
VMD (background) amplitude, 
${\cal A}(\rho\rightarrow\pi^0\pi^0\gamma)_{\omega}$, with the (signal) amplitudes obtained 
before, ${\cal A}(\rho\rightarrow\pi^0\pi^0\gamma)_{\chi}$, 
${\cal A}(\rho\rightarrow\pi^0\pi^0\gamma)_{\mbox{\scriptsize L$\sigma$M}}$ or 
${\cal A}(\rho\rightarrow\pi^0\pi^0\gamma)_{\mbox{\scriptsize $\sigma$-phen}}$, is
found to be constructive in the whole kinematical region in the three cases and one 
globally has 
\begin{equation}
\label{GchiralneutralALL}
\begin{array}{ll}
\Gamma_{\rho\rightarrow\pi^0\pi^0\gamma}^{\chi+\omega}=4.40\ \mbox{keV}\ ,
& B_{\rho\rightarrow\pi^0\pi^0\gamma}^{\chi+\omega}=2.95\times 10^{-5}\ ,\\[1ex]
\Gamma_{\rho\rightarrow\pi^0\pi^0\gamma}^{\mbox{\scriptsize L$\sigma$M+$\omega$}}=
6.29\ \mbox{keV}\ ,
& B_{\rho\rightarrow\pi^0\pi^0\gamma}^{\mbox{\scriptsize L$\sigma$M+$\omega$}}=
4.21\times 10^{-5}\ ,\\[1ex]
\Gamma_{\rho\rightarrow\pi^0\pi^0\gamma}^{\mbox{\scriptsize $\sigma$-phen +$\omega$}}=
5.10\ \mbox{keV}\ ,
& B_{\rho\rightarrow\pi^0\pi^0\gamma}^{\mbox{\scriptsize $\sigma$-phen +$\omega$}}=
3.42\times 10^{-5}\ .
\end{array}
\end{equation}
The corresponding spectra have been plotted in Fig.~\ref{plotneutral} from which the effects
of the $\sigma$ formation and their dependence on the $\sigma$ width can be observed.
The fact that our various signal amplitudes are important compared with the background
contribution and that their interferences are positive makes this
$\rho\rightarrow\pi^0\pi^0\gamma$ decay mostly appropriate to reveal the
$\sigma$ meson effects \cite{Escribano:2002aj}. 

On the experimental side, the SND Collaboration has reported very recently a new
measurement of the $\rho^0\rightarrow\pi^0\pi^0\gamma$ decay.
For the branching ratio, they obtain \cite{Achasov:2002jv}
\begin{equation} 
\label{SNDnew}  
B(\rho\rightarrow\pi^0\pi^0\gamma)=(4.1^{+1.0}_{-0.9}\pm 0.3)\times 10^{-5}\ ,
\end{equation} 
and therefore $\Gamma(\rho\rightarrow\pi^0\pi^0\gamma)=(6.1^{+1.6}_{-1.4})$ keV.
This new value is in agreement with the first measurement \cite{Achasov:2000zr}
\begin{equation} 
\label{SNDold}  
B(\rho\rightarrow\pi^0\pi^0\gamma)=(4.8^{+3.4}_{-1.8}\pm 0.2)\times 10^{-5}\ .
\end{equation}
Comparison with our predictions indicates that a substantial $\sigma$ meson contribution
is needed. Unfortunately, more crucial data on the $\pi^0\pi^0$ invariant mass spectrum have not
been reported yet. 

\section{$\rho^0\rightarrow\pi^+\pi^-\gamma$}
The background amplitude for this charged decay mode is more involved than for the previous,
neutral case. Apart from the additional amplitude, 
${\cal A}(\rho\rightarrow\pi^+\pi^-\gamma)_{\mbox{\scriptsize non-$\sigma$}}$,
generated by the second term in Eq.~(\ref{A4PchiralchargedlF}),
{\it i.e.}~the term not linked to $\sigma$ formation in the $s$-channel, we have further 
$t$- and $u$-channel contributions.
The dominant one, particularly for low photon energies, is the bremsstrahlung
amplitude, ${\cal A}(\rho\rightarrow\pi^+\pi^-\gamma)_{\mbox{\scriptsize brems}}$,
while the other one originates from $a_1 (1260)$ contributions. They can be 
regarded as $J^{PC}= 0^{-+}$ and $1^{++}$ exchanges in the $t$ and $u$ channels. 
Both contributions can be related to tree-level amplitudes and to the set of one-loop
diagrams which are specific for $\gamma\gamma\rightarrow\pi^+\pi^-$ in the analysis of
Ref.~\cite{Bijnens:1987dc}. Contrasting with the previously discussed chiral-loop diagrams
---which contributed to both the neutral and charged decay channels with finite corrections---
this new set includes divergent vertex corrections and mass insertions.
One of these divergences, appearing only in our case with $(q^*)^2 = m^2_\rho \neq 0$,
requires the contribution of a $\rho$-dominated counterterm which leads to an amplitude,  
${\cal A}(\rho\rightarrow\pi^+\pi^-\gamma)_{\mbox{\scriptsize brems}}$, including the one-loop
effects in the physical value of the $\rho\pi\pi$ coupling constant $g$ (those for the real photon
vanish with $q^2 =0$). The other piece requires the term in Ref.~\cite{Bijnens:1987dc} 
containing the combination of low-energy constants $L^r_9 +L^r_{10}$ which is known to be
saturated by pure axial resonance exchange \cite{Ecker:1988te}. It thus generates an $a_1$(1260)
contribution, 
${\cal A}(\rho\rightarrow\pi^+\pi^-\gamma)_{a_1} = 
16 \sqrt{2} ge ( L^r_9 +L^r_{10}) \{a\}$, which for $L^r_9 +L^r_{10} \simeq 1.4 \times 10^{-3}$ is
well below our signal amplitude, and can safely be neglected. 

The remaining, bremsstrahlung contribution is
well known \cite{Singer}--\cite{Huber:1995bu} 
\begin{equation} 
\label{Abremss}
\begin{array}{l}
{\cal A}(\rho\rightarrow\pi^+\pi^-\gamma)_{\mbox{\scriptsize brems}}=
2\sqrt{2}eg\\[1ex]
\quad\times
\left[\epsilon^\ast\cdot\epsilon
-\frac{1}{2}\left(\frac{\epsilon\cdot p_{+}}{q\cdot p_{+}}+
                  \frac{\epsilon\cdot p_{-}}{q\cdot p_{-}}\right)
            \epsilon^\ast\cdot q
-\frac{1}{2}\left(\frac{\epsilon\cdot p_{+}}{q\cdot p_{+}}-
                  \frac{\epsilon\cdot p_{-}}{q\cdot p_{-}}\right)
	    \epsilon^\ast\cdot (p_{+}-p_{-})\right]\ ,
\end{array}
\end{equation}  
and, by itself, it leads to 
\begin{equation} 
\label{Gbremss} 
\Gamma(\rho\rightarrow\pi^+\pi^-\gamma)_{\mbox{\scriptsize brems}}=1.706\ \mbox{MeV}
\quad\mbox{for $E_{\gamma}>50$ MeV}\ .
\end{equation}
\FIGURE{
\centerline{\includegraphics[width=0.85\textwidth]{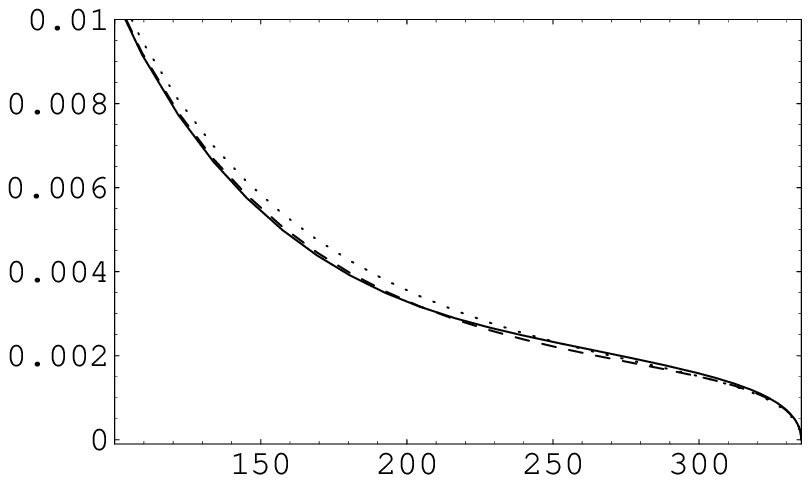}}  
\caption{\small 
$d\Gamma(\rho\rightarrow\pi^+\pi^-\gamma)/dE_{\gamma}$
as a function of the photon energy $E_{\gamma}$ (MeV) for 
$E_{\gamma}> 100$ MeV. 
The various predictions are for the input values:  
$m_\sigma = 500$ MeV and $\Gamma_\sigma = 300$ MeV (solid line);   
$m_\sigma = 500$ MeV and $\Gamma_\sigma = 500$ MeV (dashed line).
The chiral-loop prediction with no scalars is also included for comparison
(dotted line).} 
\label{plotcharged} 
}

The various signal amplitudes, ${\cal A}_\sigma (s)$, have to be added to that from background, 
${\cal A}_{\mbox{\scriptsize backg}}\equiv 
 {\cal A}_{\mbox{\scriptsize brems}}+{\cal A}_{\mbox{\scriptsize non-$\sigma$}}+
 {\cal A}_{a_1}\simeq 
 {\cal A}_{\mbox{\scriptsize brems}}+{\cal A}_{\mbox{\scriptsize non-$\sigma$}}$.
This leads to the predictions displayed in Fig.~\ref{plotcharged}  
for $E_\gamma > 100$ MeV, where  $\sigma$ meson exchange effects are visible. 
These are moderately dependent on the $\sigma$ width but show a
depletion of events below $E_\gamma \simeq 250$ MeV when compared to the chiral amplitude.  
The integrated results for $E_{\gamma}>50$ MeV, being dominated by bremsstrahlung at 
low $E_\gamma$, are less interesting but included for completeness 
\begin{equation}
\label{GchiralchargedALL} 
\begin{array}{ll} 
\Gamma_{\rho\rightarrow\pi^+\pi^-\gamma}^{\mbox{\scriptsize $\chi$+backg}}=
1.748\ \mbox{MeV}\ ,
& B_{\rho\rightarrow\pi^+\pi^-\gamma}^{\mbox{\scriptsize $\chi$+backg}}=
1.171\times 10^{-2}\ ,\\[1ex]
\Gamma_{\rho\rightarrow\pi^+\pi^-\gamma}^{\mbox{\scriptsize L$\sigma$M+backg}}=
1.698\ \mbox{MeV}\ ,
& B_{\rho\rightarrow\pi^+\pi^-\gamma}^{\mbox{\scriptsize L$\sigma$M+backg}}=
1.138\times 10^{-2}\ ,\\[1ex]
\Gamma_{\rho\rightarrow\pi^+\pi^-\gamma}^{\mbox{\scriptsize $\sigma$-phen+backg}}=
1.696\ \mbox{MeV}\ ,
& B_{\rho\rightarrow\pi^+\pi^-\gamma}^{\mbox{\scriptsize $\sigma$-phen+backg}}=
1.136\times 10^{-2}\ .
\end{array}
\end{equation}

For the $\rho^0\rightarrow\pi^+\pi^-\gamma$ decay, the present experimental
branching ratio is \cite{Dolinsky:vq,Vasserman:yr}
\begin{equation} 
\label{bremss}  
B(\rho\rightarrow\pi^+\pi^+\gamma)=(0.99\pm 0.04\pm 0.15)\%
\quad\mbox{for $E_{\gamma}>50$ MeV}\ ,
\end{equation}
quite compatible with all our results. 
The observed photon spectrum compares rather favorably with pure 
bremsstrahlung emission except (possibly) for the last bin,  
where the $\sigma$ amplitudes moderately contribute to improve the agreement.  
Finally, a model-independent upper limit of the branching ratio of the
$\rho^0\rightarrow\pi^+\pi^-\gamma$ decay via scalar resonance exchange 
was found to be
$B(\rho\rightarrow\pi^+\pi^+\gamma)< 5\times 10^{-3}\ \mbox{(90\% CL)}$
\cite{Dolinsky:vq,Vasserman:yr}
and thus fully respected in our approach. 

\section{Comments and conclusions}
Apart from old attempts to identify $\sigma$ meson contributions to 
$\rho^0\rightarrow\pi\pi\gamma$ decays \cite{renard}, other authors have 
reconsidered the issue more recently.  
Oset and collaborators \cite{Marco:1999df,Palomar:2001vg}, 
for instance, have discussed these processes in their unitarized 
chiral-loop approach where the $\sigma$ meson pole is dynamically 
generated; this makes their approach, as well as their results, quite 
different from ours. The same happens with another series of papers 
by Gokalp {\it et al.}~\cite{Gokalp:2000ir}--\cite{Gokalp:2003uf}, 
where $\sigma$ meson effects are {\it added} to the chiral-loop 
contribution; in this way, an attractive feature of our treatment, 
namely, that in the limit of high $m_\sigma$ one recovers the 
expected and well defined chiral-loop amplitude, is lost. 

In conclusion,
$\rho^0\rightarrow\pi\pi\gamma$ decays have been shown to be an important source of
information on the low-mass $\pi \pi$ spectrum in the $s$-channel. A 
global analysis of both processes, with a common amplitude interfering 
with markedly different but well stablished backgrounds, should 
contribute to clarify the $\sigma$ meson status. According to our 
analysis, present data already suggest the existence of such a 
low-mass state. Moderately improved data on $\rho^0\rightarrow\pi\pi\gamma$ 
decays could be decisive to settle the issue. 

\acknowledgments
Work partly supported by the EU, HPRN-CT-2002-00311, EURIDICE network, and the 
Ministerio de Ciencia y Tecnolog\'{\i}a and FEDER, FPA2002-00748EU.
R.~E.~acknowledges D.~Black, J.~A.~Oller and J.~R~Pel\'aez for providing us with 
the different sets of $I=J=0$ $\pi\pi$ phaseshift experimental data.


\end{document}